\documentclass[aps,pre,preprint,groupedaddress]{revtex4}

\usepackage{graphicx}
\usepackage{amsmath,amsthm,amsfonts, latexsym}

\begin{document}

\def\be{\begin{equation}}
\def\ee{\end{equation}}
\def\bea{\begin{eqnarray}}
\def\eea{\end{eqnarray}}
\newcommand{\avg}[1]{\langle{#1}\rangle}
\newcommand{\Avg}[1]{\left\langle{#1}\right\rangle}


\title{Tsallis mapping in growing complex networks with fitness}


\author{Guifeng Su$^{1}$,
Xiaobing Zhang$^{2}$,
Yi Zhang$^{1}$}
\email{yizhang@shnu.edu.cn}
\affiliation{$^{1}$ Department of Physics, Shanghai Normal University, Shanghai 200230, People's Republic of China \\
$^{2}$ School of Physics, Nankai University, Tianjin 300371,
People's Republic of China}



\begin{abstract}
We introduce Tsallis mapping in Bianconi-Barab\'asi (B-B) fitness
model of growing networks. This mapping addresses the dynamical
behavior of the fitness model within the framework of nonextensive
statistics mechanics, which is characterized by a dimensionless
nonextensivity parameter $q$. It is found that this new
phenomenological parameter plays an important role in the evolution
of networks: the underlying evolving networks may undergo a
different phases depending on the $q$ exponents, comparing to the
original B-B fitness model, and the corresponding critical transition
temperature $T_C$ is also identified.
\end{abstract}


\maketitle


\section{Introduction}
\label{sec1}

The seminal work by Watts and Strogatz on small-world
networks~\cite{Watts98}, and that by Barab\'asi and Albert on
scale-free networks (i.e., B-A networks)~\cite{BA}, have
pioneered a wave of research activities on the complex networks
in last decade (see, for instance, recent review
articles~\cite{Strogatz01, Albert02, Reviews},
books~\cite{Books, Doro_bks}, and references therein). Research
toward understanding the new critical phenomena in complex
networks, e.g., the structural phase transitions in networks,
is also a rapidly developing area~\cite{Doro_rev}. It is found
that the condensation phase
transitions occur in many complex network models~\cite{Evans,
fitness, Burda, Godreche01, Godreche05, Mezard, Krapivsky, BEC,
Rodgers02, Doro03, Borgs07, Noh, Gergely04, Biely06}. Particularly, 
in Ref.~\cite{BEC},
Bianconi-Barab\'asi (B-B) fitness model~\cite{fitness} was mapped
into an equilibrium Bose gas, a Bose-Einstein condensation (BEC)
behavior was obtained, and the critical point for condensation was
identified by mean-field arguments. There the well-known B-A
networks~\cite{BA} are generalized by introducing the {\it fitness}
parameter, $\eta_i$. At each time step, a new node with $m$ links is
added to the nodes that already presented in the network. The
fitness $\eta_i$, accounting for the competition for links in the
network, is chosen randomly from a distribution. The probability
$\Pi_i$ that a new node connects one of its $m$ links to a node $i$
depends on the number of links, $k_i(t)$, and on its fitness
$\eta_i$, is
\be
\Pi_i = \frac{\eta_i k_i}{\sum_j \eta_j k_j}.
\label{Pi}
\ee

In order to map the fitness model into Bose gases, one assigns an
``energy'' $\varepsilon_i$, introduced randomly from a distribution
$g(\varepsilon)$, for the $i$th node, and the fitness of the node,
$\eta_i$, is defined as \be \eta_i = e^{-\beta \varepsilon_i}, \ee
where the parameter $\beta$, can be identified as inverse
``temperature'' $\beta = 1/T$. This growing network can be solved
in a mean-field approximation and a ``chemical potential'' $\mu$
is determined by the self-consistent equation
\be I(\beta, \mu) = \int d \varepsilon g(\varepsilon) n(\varepsilon)
= 1 , \label{selfc} \ee where the Bose occupation number
$n(\varepsilon) = 1/(e^{\beta(\varepsilon - \mu)} - 1)$ is the
number of links attached by the preferential attachment mechanism
to nodes with ``energy'' $\varepsilon$. A BEC phase transition
occurs in the networks if $I(\beta, 0) < 1$, i.e., the
self-consistent equation (\ref{selfc}) has no solution (in such case
the self-consistent approach fails). A necessary condition for
condensation is that the distribution $g(\varepsilon) \rightarrow 0$
for $\varepsilon \rightarrow 0$. For this type distributions there
exists a critical temperature $T_c = 1/\beta_c$ such that $I(\beta,
0) < 1$ for $T < T_c$.

However, it should be noted that BE statistics appearing in B-B
growing network is only formally equivalent to that in ideal Bose
gases. This is due to the fact that the evolution of B-B network is
in essential irreversible and out-of-equilibrium, while the BEC of
ideal Bose gases is a temperature-driven equilibrium process. On the
other hand, it was argued~\cite{Song05} a wide variety of complex
networks have self-similar, fractal property. Under such
circumstances, a particularly useful and appropriate framework to
describe the dynamical evolution of the network is so-called Tsallis
nonextensive statistical mechanics~\cite{Tsallis88, Tsallis90s,
Salinas99}. It is a certain generalization of the usual
Boltzmann-Gibbs (BG) formulation. For the applications of the
formulation in many fields, including complex networks, one may see,
e.g., Refs.~\cite{TsallisApp, Soarse04} and references therein. In
present work, we consider the possibility of embedding the B-B
fitness model in the framework of Tsallis formulation of the
non-extensive statistical mechanics. The paper is organized as
follows: in Sec.~\ref{sec2}, we briefly review Tsallis formulation
of the nonextensive statistical mechanics, and introduce Tsallis
mapping in the B-B fitness model of the complex network. We show
that the corresponding rate equation has a power-law solution under
some approximation within the high temperature regime, in
thermodynamic limit. For general situations without the
thermodynamic limit, the numerical simulations are need. The main
simulation results are shown and discussed in Sec.~\ref{sec3}.
Finally, we summarize and draw our conclusions.

\section{Tsallis mapping in B-B fitness networks}
\label{sec2}

In Tsallis formulation of non-extensive statistical mechanics, the
central equation is so-called the nonextensive entropy, which is
defined as

\be
S_q = \frac{1}{q-1}  (1 - \sum_i p^q_i) ,
\ee

where $p_i$ is the probabilities for the ensemble to be in the state
$i$. The non-extensive parameter $q$ describes the degree of
deviations of a physical quantity from extensive and therefore may be
regarded as a new phenomenological parameter of the model used,
different values of $q - 1$ quantifies the departure from the BG limit.
The standard results of BG statistical mechanics are recovered in the
limit $q \rightarrow 1$. The application of the non-extensive
statistical mechanics to Boson gases predicts a $q$-deformed BE
statistics~\cite{Tsallis90s}:
\be
n_q (\varepsilon) = \frac{1}{e_{q}^{\beta (\varepsilon - \mu)} - 1},
\label{qBE}
\ee
where $n_q (\varepsilon)$ is the occupation number of an energy level
$\varepsilon$ and $\mu$ the chemical potential. The parameter $\beta$
is the inverse temperature. A $q$-exponential function is introduced
in Eqn. (\ref{qBE}) as follows~\cite{Tsallis94}:
\be
e^{z}_q \equiv [1 + (1 - q)z]^{1/(1-q)} ,
\ee
if the argument $1 + (1 - q)z$ is positive,
$e^{z}_q $ equals to zero otherwise. Its inverse function, the
$q$-logarithm, for any positive real number $z \in \mathbb{R}^{+}$,
is
\be
\ln_q z \equiv \frac{z^{1 - q} - 1}{1 - q} .
\label{qlog}
\ee
Both $q$-logarithm and $q$-exponential functions go back to the
usual logarithm and exponential functions, respectively, when one
takes $q \rightarrow 1$ limit.

In order to map B-B fitness model to a boson gas with Tsallis
statistics, under the self-consistency condition, we assign the
$i$th node with a fitness, $\eta_i$, by
\be
\eta_i = e^{-\beta \varepsilon_i}_q
       = \Big(1 + (1 - q)(-\beta \varepsilon_i)\Big)^{1/(1 - q)} ,
       \ee
and its inverse is
\be
\varepsilon_i = -\frac{1}{\beta} \ln_q\eta_i
              = -\frac{1}{\beta}\frac{\eta_i^{1 - q} - 1}{1 - q} ,
\ee
where $\varepsilon_i$ follows the energy level distribution
$g(\varepsilon) = C \varepsilon^{\theta}$ and $C$ is a
normalization factor. A necessary condition for condensation is that
when $\varepsilon \rightarrow 0$, $g(\varepsilon) \rightarrow 0$
consistently. The resulting rate equation which describes the degree
of connectivity at time $t_i$ of the $i$th node (energy level),
$k_i(\varepsilon_i, t, t_i)$, is then:
\be
\label{rate}
\frac{\partial k_i(\varepsilon_i, t, t_i)}{\partial t} =
\frac{e^{-\beta \varepsilon_i}_{q} k_i (\varepsilon_i, t, t_i)}{t
Z^q_{t}} , \ee
where
\be
\label{zq}
Z^{q}_{t} = \frac{1}{mt} \sum^{t}_{j = 1}
e^{-\beta \varepsilon_j}_{q} k_j (\varepsilon_j, t, t_j) ,
\ee
is the $q$-deformed partition function. Following the standard
statistical mechanics~\cite{KHuang}, the definition of
fugacity $z$ is given, in a consistent way, by $q$-deformed form,
$z^{-1} = e^{-\beta \mu}_q$, in which $\mu$ is the chemical potential.

The phase structure of this B-B fitness model with Tsallis mapping
is now controlled by two parameters, i.e., the inverse temperature
$\beta$ and the nonextensive parameter $q$, instead of (only) one
parameter $\beta$ in the original B-B fitness model. These two
parameters both affect the phase transition of the network. A few
limiting situations are worthy of being discussed before further
detailed numerical simulations. First, in $q \rightarrow 1$ limit,
the B-B model with Tsallis mapping goes back to the original one
with Bose gas mapping, due to the fact that Tsallis $q$-deformed
expressions approach to the usual ones, and all the known results of
the B-B model are recovered as they should. Henceforth, the
scale-free phase in original B-B fitness model is trivially
recovered when all nodes have the same fitness, i.e.,
$g(\varepsilon) \sim \delta (\varepsilon)$~\cite{BA, BEC}. Second,
with $q \rightarrow 1$ limit and the thermodynamic limit $t
\rightarrow \infty$ as well, we find that the rate equation
(\ref{rate}) has a power law solution under some appropriate
approximation, within high temperature region, in which $\beta$ is a
small quantity. This phase, as matter of fact, is similar to the
fit-get-rich (FGR) phase in the original B-B fitness
model~\cite{fitness}.

To see the latter, we assume that the solution takes the following
form:
\be
\label{powerlaw}
k_i(\varepsilon_i, t, t_i) \sim
\Big(\frac{t}{t_i}\Big)^{f_q(\varepsilon_i)} ,
\ee
where
$f_q(\varepsilon)$ is some unknown power-law exponent and
to
be determined later. For simplicity we will write
$f_q(\varepsilon) \equiv f_q$ from now on. Taking the energy
distribution of the nodes $g(\varepsilon)$ into account, the
partition function becomes
\be
\overline{Z^{q}_{t}} \sim  \int d \varepsilon g(\varepsilon)
\int^{t}_{1} \frac{dt_0}{t} e^{-\beta \varepsilon}_{q} 
k_i(\varepsilon, t, t_0),
\ee
here we use
$\overline{Z^{q}_{t}}$ stand for the partition function averaging
over $g(\varepsilon)$. Now inserting $k_i$ (in Eqn.(\ref{powerlaw}))
into $\overline{Z^{q}_{t}}$, one obtains
\bea
\overline{Z^{q}_{t}}
& \sim & \int d \varepsilon g(\varepsilon) \int^{t}_{1} \frac{d t_0}{t}
e^{-\beta \varepsilon}_{q} \Big(\frac{t}{t_0} \Big)^{f_q} \nonumber \\
& = & \int d \varepsilon g(\varepsilon) e^{-\beta \varepsilon}_{q}
t^{f_q - 1} \bigg(\frac{t^{1 - f_q} - 1}{1 - f_q} \bigg) \nonumber \\
& = & \int d \varepsilon g(\varepsilon)
e^{-\beta \varepsilon}_{q} t^{f_q - 1} (\ln_{f_q} t)  \nonumber \\
& \sim & z^{-1} (1 - t^{-(1 - f_q)}) , \label{last}
\label{Zqt}
\eea
where $\ln_{f_q} t$ is the $f_q$-logarithm as introduced in
Eqn. (\ref{qlog}) and the inverse fugacity $z^{-1}$ is
\be
\label{invfugacity}
z^{-1} = \int d \varepsilon g(\varepsilon)
         \frac{e^{-\beta \varepsilon}_q}{1 - f_q} .
\ee
Note that in Eqn. (\ref{Zqt}), $1 - f_q > 0$, such that in
thermodynamic limit $t \rightarrow \infty$, $t^{-(1 - f_q)}
\rightarrow 0$, and hence $\overline{Z^{q}_{t}} \sim z^{-1}$
in the corresponding limit. This makes one able to determine
aforementioned power-law exponent $f_q$. Note that in high
temperature region, $\beta$ is small, and that $q$ is very close
to unity (such that $q-1$ is small as well), so we can take
the approximation
$e^{-\beta \varepsilon}_q/e^{-\beta \mu}_q \approx
e^{-\beta(\varepsilon-\mu)}_q$~\cite{error},
as a result, the exponent $f_q$ of the power-law solution in
Eqn. (\ref{powerlaw}) is,
\be
\label{fq}
f_q = e^{-\beta (\varepsilon - \mu)}_q ,
\ee
in $q \rightarrow 1$ limit, this exponent goes back to the
one in the original B-B fitness model~\cite{BEC}.
The fugacity becomes
\be
\label{limit}
z^{-1} = \lim_{t \rightarrow \infty} \overline{Z^{q}_{t}}
       = \int d\varepsilon g(\varepsilon)
         \frac{e^{-\beta \varepsilon}_q}{1 - f_q(\varepsilon)} ,
\ee
and the chemical potential $\mu$ is the solution of the following
equation,
\be
\label{Iq}
I_q(\beta, \mu) = \int d \varepsilon g(\varepsilon)
\frac{1}{e^{\beta (\varepsilon - \mu)}_q - 1} = 1 ,
\ee
which indicates a mapping to Bose gas with Tsallis statistics,
and the corresponding occupation number of a energy level
$\varepsilon$ is
$n(\varepsilon) = (e^{\beta (\varepsilon-\mu)}_q - 1)^{-1}$.
This just follows Tsallis statistics.

For general situations without taking into account above limits, one
has to resort to numerical simulations for current model with
Tsallis mapping.

\section{Numerical simulation results and discussions}
\label{sec3}

In the original B-B fitness model, the signal of the
condensation phase transition lies on the change of sign of the
chemical potential $\mu$, i.e., when $\mu$ experiences a change
from negative value to positive value within some regime, that
change identifies the critical point of the corresponding phase
transition. Similarly, we use the same criterion for the phase
transition in the B-B fitness network with Tsallis mapping.
This is done by numerically computation of the chemical
potential $\mu$, without the thermodynamic limit,
according to the following way:
\be
\mu = -\frac{1}{\beta} \ln_q \overline{Z^{q}_{t}}
    = -\frac{1}{\beta} \frac{(\overline{Z^{q}_{t}})^{1 - q} - 1}{1 - q} ,
    \label{mu_q}
\ee
where $\ln_q \overline{Z^{q}_{t}}$ is the $q$-logarithm of
$\overline{Z^{q}_{t}}$ which is defined in Eqn. (\ref{zq}).
Note that in above numerical simulations, the computation of
$\mu$ strictly follows its definition and has no approximations in
them.

\begin{figure}
\includegraphics[width=8cm]{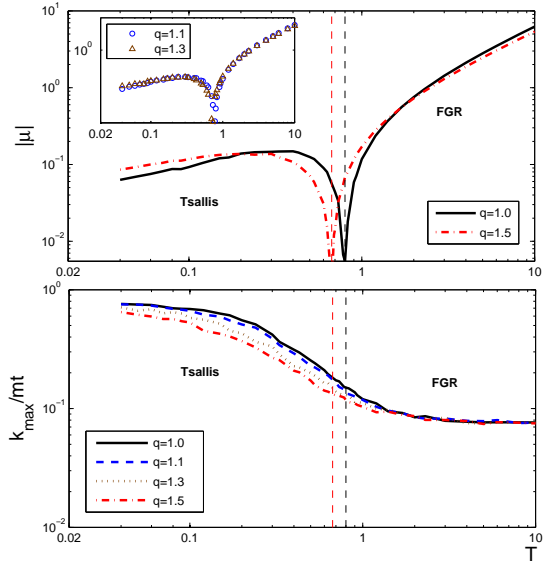}\\
\caption{Numerical simulations of the fitness model with Tsallis
mapping for growing complex networks. An energy level distribution
$g(\varepsilon) \sim \varepsilon$ is taken in the figure. We
set evolving times (system size) to be $t = 10^3$, and averaged over
$300$ runs. Upper Panel: Numerically calculated chemical potential
using the energies distribution with $\varepsilon \in$ (0, 1). The
temperatures at which $\mu$ changes sign correspond to the sharp
drop in $|\mu|$ on the figure, and identify the different critical
temperatures $T^{q}_c$ (two vertical dashed lines in the figure) for
Tsallis condensation phase with different $q$-value: $q = 1.0$
(solid line), corresponds to the results of B-B fitness model; $q =
1.5$ (dot-dashed line). The inset plot shows the chemical potential
for another choices of $q$ in between: $q = 1.1$ (open circles), and
$q = 1.3$ (open triangles), respectively. Lower Panel: Fraction of
the total number of links connected to the most connected (or
``winner") node, $k_{max}/(mt)$, plotted as a function of T, shown
for $m = 2$ and for different $q$-s: $q = 1.0$ (solid line), $q =
1.1$ (dashed line), $q = 1.3$ (dotted line), and $q = 1.5$
(dot-dashed line). \label{fig1}}
\end{figure}

\begin{figure}
\includegraphics[width=8cm]{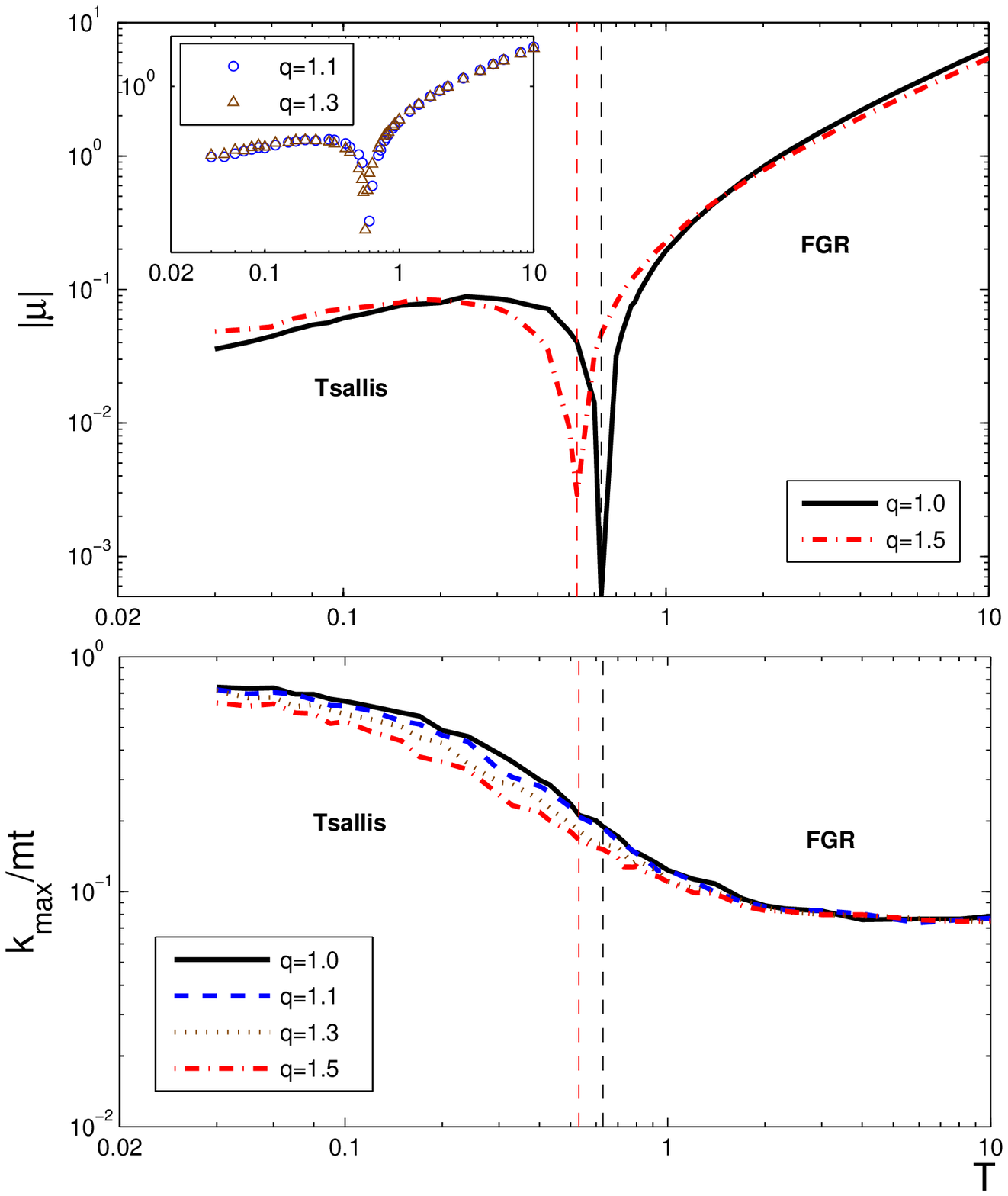}\\
\caption{Similar to Fig.~\ref{fig1}, but simulated for energy level
distribution $g(\varepsilon) \sim \sqrt{\varepsilon}$ and with
different $q$-value. Upper Panel: the chemical potential $\mu$ vs.
temperature $T$ with $q = 1.0$ (solid line), and $q = 1.5$
(dot-dashed line). The inset plot shows $q = 1.1$ (open circles),
and $q = 1.3$ (open triangles), respectively. Lower Panel: the share
of links, $k_{max}/(mt)$ plotted as a function of $T$, shown for
$m = 2$ and for different $q$-value: $q = 1.0$ (solid line),
$q = 1.1$ (dashed line), $q = 1.3$ (dotted line), and $q = 1.5$
(dot-dashed line), respectively. \label{fig2}}
\end{figure}

\begin{figure}
\includegraphics[width=8cm]{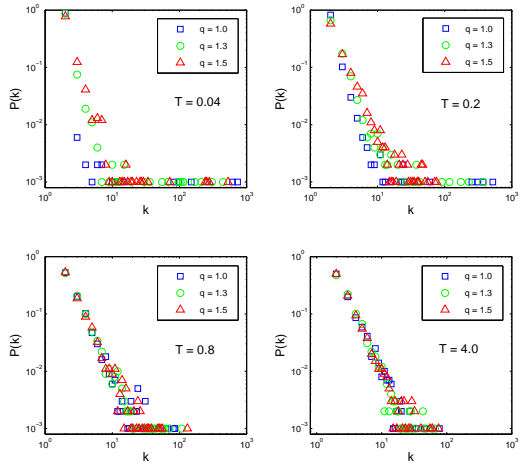}\\
\caption{The probability distribution of the degree of connectivity
$P(k)$ with $q =$ 1.0, 1.3, and 1.5, for energy level distribution
$g(\varepsilon) \sim \varepsilon$, at the different
temperatures (from low to high). Upper-left Panel: at temperature $T
= 0.04$, and $q = 1.0$ (open sqaures), $q = 1.3$ (open circles), and
$q = 1.5$ (open triangles); Upper-right Panel: similar to the
upper-left panel but at temperature $T = 0.2$; Lower-left and -right
Panels: similar to upper panels but at temperatures $T = 0.8$ and $T
= 4.0$, respectively. \label{fig3}}
\end{figure}

The ``phase diagrams" of the B-B fitness network with Tsallis mapping
are demonstrated by our numerical simulations and are shown in
Fig.~\ref{fig1} --~\ref{fig3}. In both figures, the size of the
evolving network, or the time steps, is taken to be $t = 10^3$ and
we fix $m = 2$. In Fig.~\ref{fig1}, the absolute value of the
chemical potential $\mu$, which is defined in Eqn. (\ref{mu_q}), is
numerically computed with an average over $300$ runs, using the
energy level distribution $g(\varepsilon) = 2 \varepsilon$ for
energies $\varepsilon \in $ (0, 1), i.e., $\theta = 1$ and $C = 2$,
and for $q = 1.0$, $1.1$, $1.3$, and $1.5$, respectively. As
expected, for $q = 1.0$, the phases emerge in the system are
coincident with those in the original B-B fitness model: the
FGR phase and the BEC phase (the solid line in Fig.~\ref{fig1}),
and the same transition temperature is located at $T_{BE} \sim 0.8$
(the vertical dash-line on the right).
However, when the value of
parameter $q$ deviates from $1$, e.g., for $q = 1.1$, $1.3$, and/or
$1.5$, the phases correspondingly change depending on both the
specific value of $q$ and the temperature they take (see the
dot-dashed line in Fig.~\ref{fig1}). Also, it can be seen from the
figure that the transition temperature $T^q_c$ decreases and
$T^q_c \sim 0.67$ (the vertical dashed line on the left in
Fig.~\ref{fig1}). In addition, in $T < T^q_c$ regime, the whole
``Tsallis condensate" curve deviates from the BEC curve. On
the other hand, it is no surprise that the effect of $q$ parameter
disappears gradually in the high temperature limit. The effect of
different energy level distribution is shown in Fig.~\ref{fig2},
with $\theta = 1/2$. We also plot the fraction of the total number
of links connected to the most connected (or ``winner") node,
$k_{max}/mt$, as a function of $T$ (lower panel in Fig.~\ref{fig1}
and Fig.~\ref{fig2}).

We also plot in Fig. .\ref{fig3} the connectivity degree distribution 
of the B-B fitness network with Tsallis mapping, i.e., the probability
distribution $P(k)$ vs. the connectivity degree $k$, for different 
values of parameter $q$ and temperatures $T$. In the figure, $P(k)$ 
is plotted for $g(\varepsilon) \sim \varepsilon$, with nonextensive
parameter $q = 1.0$, $1.3$, and $1.5$, and with temperature $T=0.08$, 
$T=0.4$, $T=0.8$, and $T=4.0$, respectively.
Since the critical temperature is $T_C=0.67$, in the latter 
cases (i.e., $T=0.8$ and $T=4.0$), the temperature $T > T_C$. 
As we expected, the Tsallis mapping gives the power-law degree 
distribution for $T > T_C$, $P(k) \sim k^{-\gamma}$, with 
$\gamma \sim f_q$ (see the lower panels in Fig.\ref{fig3}, in 
which $T=0.8$ and $T=4.0$, respectively). Note that, first, the 
nonextensive parameter $q$ controls the slope of $P(k)$ at given 
temperature; second, data points with $q=1$ (open squares with 
blue color in Fig.~\ref{fig3}) in fact correspond to the FGR 
phase in the original B-B model for $T > T_C$. When the temperature 
decreases, for example, at the critical point $T_C \approx 0.67$, 
the chemical potential changes its sign from positive to negative, 
and this correspondes to the situation that a condensate phase 
starts to emerge. This means the network experiencs a topological 
transition. In the whole condensate phase region (or $T < T_C$),
the power-law solution of the degree distribution hold in FGR region 
breaks down and hence $P(k)$ deviates from power-law distribution. 
However, it is still controlled by the nonextensive parameter $q$, 
as we can clearly see from the figure.

\section{Conclusions}
\label{sec4}

In conclusion, we consider Bianconi-Barab\'asi fitness model of
growing networks in the framework of non-extensive statistical
mechanics via Tsallis mapping in the fitness parameter $\eta_i$. 
We find that the phenomenological nonextensive parameter $q$ plays 
an important role in the dynamical evolution of complex networks: 
the underlying evolving networks may undergo some FGR-like phase 
in high temperature region with a nonextensive parameter controlled 
power-law degree distribution and it goes back to the original FGR 
phase of B-B fitness model in $q \rightarrow 1$ limit; on the other 
hand, in low temperature regime, particularly for $T < T_C$ (where 
$T_C$ is the critical temperature), a condensate phase emerges. For 
this condensate phase, its connectivity degree distribution $P(k)$ 
deviates from the power-law behavior due to the fact that the 
power-law solution hold in FGR region breaks down, but its specific 
form is controlled by the nonextensive parameter $q$. The 
corresponding critical transition temperature $T_C$ is also 
identified and found decrease comparing to that of the original B-B 
fitness model.

Tsallis formulation of nonextensive statistical mechanics not only
provides a natural framework for describing the dynamical evolution 
of the growing complex networks, but also introduces a new complexity 
or entropy measure. This entropy measure is 
expected to be different from the traditional Shannon entropy or Gibbs 
entropy measures, e.g., those introduced in Refs.~\cite{Gfel05, Boga06, 
Bian09, Lato08, Pass08} in the context of complex networks. The current
research opens a different angle to explore such measure, and our 
results may stimulate some further investigations on this regard~\cite{SZZ11}.     

\begin{acknowledgments}
We acknowledge to the national Natural Science Foundation of China
(NSFC) for financial support, under Contract No. 10875058. 
The authors Y.Z. and G.S. also thank the support from the Initiative 
Plan of Shanghai Education Committee (Project No. 10YZ76) and the 
Scientific Research Foundation for the Returned Overseas Chinese 
Scholars, State Education Ministry (SRF for ROCS, SEM).
\end{acknowledgments}



\begin{thebibliography}{99}
\bibitem{error}
Simple numerical analysis shows that the relative error of this
approximation is between $10^{-3}$ and $10^{-1}$, for most values
of temperature in high temperature region.
\end{thebibliography}
\end{document}